\documentclass{article}

\usepackage{amsfonts,amssymb, amsmath,mathrsfs}

\usepackage[english]{babel}

\DeclareMathAlphabet{\mathpzc}{OT1}{pzc}{m}{it}

\def\on#1#2{\mathop{\vbox{\ialign{##\crcr\noalign{\kern2pt}
$\scriptstyle{#2}$\crcr\noalign{\kern2pt\nointerlineskip}
\kern-2pt$\hfil\displaystyle{#1}\hfil$\crcr}}}\limits}

\def\nn{ \nonumber }
\def\bq{ \begin{equation} }
\def\eq{ \end{equation} }
\def\ben{ \begin{eqnarray} }
\def\en{ \end{eqnarray} }

\def\e{{\rm e}}

\newtheorem{exa}{Example}
\newenvironment{exam}{\begin{exa} \rm }{\end{exa}}

\newtheorem{re}{Remark}
\newenvironment{rem}{\begin{re} \rm }{\end{re}}

\begin{document}

%%%%%%%%%%%% TITLE %%%%%%%%%%%%%%

\title{On maximally  superintegrable  systems.}

\author{A V Tsiganov \\
\it\small
St.Petersburg State University, St.Petersburg, Russia\\
\it\small e--mail: tsiganov@mph.phys.spbu.ru}
\date{}
 \maketitle

\begin{abstract}
Locally any completely integrable system is maximally superintegrable  system such as we have the necessary number of the action-angle variables. The main problem is the construction of the single-valued additional integrals of motion  on the whole phase space by using  these multi-valued action-angle variables. Some constructions of the additional integrals of motion for the St\"ackel systems and for the integrable systems related with two different quadratic $r$-matrix algebras are discussed. Among these system there are the open Heisenberg magnet and the open Toda lattices associated with the different root systems.
\end{abstract}

\section{Introduction}
In classical mechanics a Hamiltonian system on a $2n$-dimensional phase space $M$ is called completely integrable in Liouville's sense if it possesses $n$ functionally independent integrals
of motion $H_1,\dots,H_n$ in the involution:
\[\dfrac{dH_i}{dt}=\{H,H_i\}=0,\qquad \{H_i, H_j\}= 0,\qquad i,j=1,\ldots,n,\]
where $H=H_1$ is the Hamilton function and $\{.,.\}$ is the Poisson bracket on $M$.

A superintegrable system is a system that is integrable in the Liouville sense and that, in addition to this, possesses more functionally independent integrals of motion than degrees of freedom. If the number of independent integrals takes the value $2n-1$, then the system is called maximally superintegrable \cite{win04}. There are three classic and well-known examples of such   systems, namely, the free particle that can be considered as trivial, the Kepler problem, and the harmonic oscillator with rational frequencies. In these cases all the orbits become closed for the case of bounded motions.

The Liouville classical theorem on completely integrable Hamiltonian systems
implies that almost all points of the manifold $M$ are covered by a system of
open toroidal domains with the action-angle coordinates $I=(I_1,\ldots,I_k)$ and $\omega=(\omega_1,\ldots,\omega_n)$ \cite{neh72}:
\bq\label{aa-br}
\{I_j,I_k\}=\{\omega_i,\omega_k\}=0,\qquad \{I_j,\omega_k\}=\delta_{ij}.
\eq
The independent integrals of motion $H_1,\ldots,H_n$ are functions upon the independent action variables $I_1,\ldots,I_n$ and the corresponding Jacobian does not equal to zero
\bq\label{h-mat}
\det\mathbf J\neq 0,\qquad\mbox{where} \qquad \mathbf J_{ij}= \dfrac{\partial H_i(I_1,\ldots,I_n)}{\partial I_j}\,.
\eq
Let us introduce $n$ functions
\bq\label{phi_k}
\phi_j=\sum_{k} \left(\mathbf J^{-1}\right)_{kj}\,\omega_k,\eq
such that
\bq\label{tr-alg}
\{H_i,\phi_j\}=\sum_{k=1}^{n} \mathbf J_{ik}\left(\mathbf J^{-1}\right)_{kj} =\delta_{ij}.
\eq
The ($n-1$) functions $\phi_2,\ldots,\phi_n$ are integrals of motion
\[
\dfrac{d\phi_j}{dt}=\{H_1,\phi_j\}=0,\qquad j=2,\ldots,n,
\]
which are functionally independent on $n$ functions $H_1(I),\ldots,H_n(I)$. So, in classical mechanics any completely integrable system is superintegrable system
in a neighborhood of any regular point of $M$. It means that the Hamiltonian $H=H_1$ has $2(n-1)$ integrals of motion $H_2,\ldots,H_n$ and $\phi_2,\ldots,\phi_n$ on any open toroidal domain.

Sometimes the action-angle variables are global variables on the whole phase space $M$ and, therefore, we have superintegrable systems on $M$. For instance,  the global action-angle variables for the open and periodic Toda lattices are discussed in  \cite{hk08}.

In generic case the action variables $\omega_k$ are multi-valued  functions on the whole phase space $M$. In this case  if we have some algebraic integral of motion $K$, it has to be some function on the action-angle variables. In this paper we discuss some possibilities to get polynomial integrals of motion from the multi-valued angle variables.

%%%%%%%%%%%%%%%%%%%%%%%%%%%%%%%%%%%%%%%%%%%%%%%%%%%%%%%%%%%%%%%%%%%%%%%%%%%%%%%%%5
%%%%%%%%%%%%%%%%%%%%%%%%%%%%%%%%%%%%%%%%%%%%%%%%%%%%%%%%%%%%%%%%%%%%%%%%%%%%%%%%%%
\section{The St\"{a}ckel systems.}
\setcounter{equation}{0}

The system associated with the name of St\"{a}ckel \cite{st95,ts99,ts99a} is a holonomic system on the phase space
$\mathbb R^{2n}$, with the canonical variables
$q=(q_1,\ldots,q_n)$ and $p=(p_1,\ldots,p_n)$:
\bq \Omega=\sum_{j=1}^n dp_j\wedge dq_j\,,\qquad
\{p_j,q_k\}=\delta_{jk}\,.\label{stw}
\eq
The nondegenerate $n\times n$ St\"{a}ckel matrix $S$, whose $j$ column depend on coordinate $q_j$ only, defines $n$ functionally independent integrals of motion
\bq
H_k=\sum_{j=1}^n ( S^{-1})_{jk}\Bigl(p_j^2+U_j(q_j)\Bigr)\,.
\label{fint}
\eq
From this definition one immediately
gets the separated relations
\bq
p_j^2=
\sum_{k=1}^n H_k S_{kj}(q_j)-U_j(q_j)\,.\label{stc}
\eq
It allows reducing the solution of the equations of motion to a problem in algebraic geometry \cite{bel97,du81,ts99}. We can regard each separated equation (\ref{stc}) as equation defining the hyperelliptic curve
\bq
\mathcal C_j:\quad
\mu_j^2=P_j(\lambda_j)\,,\label{sthc}
\eq
where $\mu_j$ and $\lambda_j$ are functions on $p_j$ and $q_j$ only and $P_j(\lambda_j)$ are polynomials.

The functionally independent action variables $I_k=H_k$ (\ref{fint}) have the canonical Poisson brackets (\ref{aa-br}) with the following angle variables
\bq\label{w-st}
\omega_i=\sum_{j=1}^n\int_{A_j} \dfrac{S_{ij}(\lambda_j)}{\sqrt{P_j(\lambda_j)\,}}\,\mathrm d\lambda_j,
\eq
which consist of integrals of first kind Abelian differentials on the hyperelliptic curves $ C_j$ (\ref{sthc}) \cite{bel97,du81,ts99,ts99a}.
For the coinciding curves we have to use different $A$-cycles on the Riemann surfaces in the sums (\ref{w-st}).

In generic case the action variables (\ref{w-st}) are the sum of the multi-valued  functions $\vartheta_{ij}$. However, if we are able to apply some addition theorem  to the calculation of $\omega_i$ (\ref{w-st})
\bq\label{add-th}
\omega_i=\sum_{j=1}^n \vartheta_{ij}(p_j,q_j)=\Theta_i \bigl(K_i\bigr),
\eq
where $\Theta_i$ is a multi-valued function on the algebraic argument $K_i(p,q)$  then one gets algebraic integrals of motion $K_i(p,q)$ such that
\[
\{H_1,\omega_i\}=\{H_1,\Theta_i \bigl(K_i\bigr)\}=\Theta_i'\cdot\{H_1,K_i\}=0.
\]
So, the addition theorems (\ref{add-th}) could help us to classify algebraically superintegrable systems and vice versa.

\begin{rem}
We can express hyperelliptic integrals  $\vartheta_{ij}$ (\ref{w-st}-\ref{add-th}) via the elementary functions for the  hyperelliptic curves $\mathcal C_j$ with genus $\mathrm=0$ only \cite{bel97,du81}. Namely, if \[\mbox{deg}P_j(\lambda)>2,\]
then variable $\omega_j$ (\ref{w-st}) are elliptic functions on $p$ and $q$.
\end{rem}

\begin{rem}
According to \cite{gts05} construction of the separated variables $(q,p)$ for the $L$-subset of the St\"ackel systems  is a pure computer problem. The software \cite{gts05} may be easily improved in order to calculate additional integrals of motion $\omega_k$ (\ref{w-st}) too.
\end{rem}

\begin{exam} \textbf{The (b) Drach system}
\par\noindent
Let us consider the St\"{a}ckel system defined by
\[
S=\left(\begin{array}{cc}q_1^2& q_2^2\\ 1& 1\end{array}\right),
\qquad U_{1,2}=\pm2\alpha-\dfrac{\beta\mp2\gamma}{q_{1,2}^2}\,,
\]
such that the separated relations (\ref{stc}) look like
\[
p_{1,2}^2 = H_1q_{1,2}^2+H_2\pm 2\alpha-\dfrac{\beta\mp 2\gamma}{q_{1,2}^2}\,.
\]
In variables $\mu_j=q_jp_j$ and $\lambda_j=q_j^2$ equations (\ref{sthc})
became the canonical equations defining two  Riemann surfaces:
\[
\mathcal C_{1,2}:\qquad \mu^2=P_{1,2}(\lambda)=
H_1\lambda^2+H_2\lambda \pm 2\alpha\lambda-\beta\pm 2\gamma.
\]
\end{exam}
The action variables $I_{1,2}=H_{1,2}$ have the canonical brackets (\ref{aa-br}) with the following  angle variables (\ref{w-st}):
\ben
\omega_1&=&\dfrac{1}{4}\left(\int^{q_1^2} \dfrac{\lambda}{\sqrt{P_1(\lambda)}}\,\mathrm d \lambda
+\int^{q_2^2} \dfrac{\lambda}{\sqrt{P_2(\lambda)}}\,\mathrm d \lambda\right)\,,\nn\\
\omega_2&=&\dfrac{1}{4}\left(\int^{q_1^2} \dfrac{1}{\sqrt{P_1(\lambda)}}\,\mathrm d \lambda
+\int^{q_2^2} \dfrac{1}{\sqrt{P_2(\lambda)}}\,\mathrm d \lambda\right)\,.\nn
\en
So, the Hamiltonian system with hamiltonian $H_1$ has two independent integrals of motion $H_2$ (\ref{fint})
and
\[
\omega_2=\dfrac{1}{4\sqrt{H_1}}\left[
\ln\left(p_1q_1+\dfrac{P^{\,\prime}_1}{2\sqrt{H_1}}\right) +\ln\left(p_2q_2+\dfrac{P^{\,\prime}_2}{2\sqrt{H_1}}\right)
\right],\]
where
\[P^{\,\prime}_{1,2}=\left.\frac{d P_{1,2}(\lambda)}{d\lambda}\right|_{\lambda=q_j^2}=
2H_1\,q_{1,2}^2+H_2\pm2\alpha\,.
\]
Using addition theorem
\bq\label{add-ln}
e^x\e^y=\e^{x+y}\qquad\mathrm{or}\qquad \ln(x_1)+\ln(x_2)=\ln\left(x_1x_2\right)
\eq
one gets
\[\omega_2=\dfrac{1}{4\sqrt{H_1}}\ln\left[
\left(p_1q_1+\dfrac{P^{\,\prime}_1}{2\sqrt{H_1}}\right) \left(p_2q_2+\dfrac{P^{\,\prime}_2}{2\sqrt{H_1}}\right)
\right]\,.
\]
So we can introduce algebraic integral of motion
\ben
\Phi_{1}(I,\omega)&=&\e^{4\sqrt{H_1}\omega_2}=\left(p_1q_1+\dfrac{P^{\,\prime}_1}{2\sqrt{H_1}}\right) \left(p_2q_2+\dfrac{P^{\,\prime}_2}{2\sqrt{H_1}}\right)=\nn\\
&=&\dfrac{4p_1p_2q_1q_2H_1+P^{\,\prime}_1P^{\,\prime}_2}{4H_1}
+\dfrac{p_1q_1P^{\,\prime}_2+p_2q_2P^{\,\prime}_1}{2\sqrt{H_1}}=\nn\\
&=&\dfrac{K_4}{4H_1}+\dfrac{K_3}{2\sqrt{H_1}}\,,\nn
\en
where $K_4$ and $K_3$ are polynomials fourth and third order in momenta
\bq\label{k3-dr}
K_3=2(p_1q_1\,P'_2+p_2q_2\, P'_1),\qquad
K_{4}=P'_1P'_2+4p_1q_1p_2q_2H_1,
\eq
such that \[\{H_1,K_3\}=\{H_1,K_4\}=0\qquad\textrm{and}\qquad\textrm \{H_2,K_3\}=K_4.\]
After the following change of variables
\[
x=\frac{(q_1-q_2)^2}{4}\,,\quad p_x=\frac{p_1-p_2}{q_1-q_2}\,,\qquad
y=\frac{(q_1+q_2)^2}{4}\,,\quad p_y=\frac{p_1+p_2}{q_1+q_2}\,,
\]
the Hamiltonian
\[
H_1=p_xp_y+\dfrac{\alpha}{\sqrt{xy}\,}+\dfrac{\beta}{(x-y)^2}+\dfrac{\gamma(x+y)}{\sqrt{xy\,}(x-y)^2}
\]
coincides with the Hamiltonian  for the one of the Drach systems (case (b) in \cite{ts00}), whereas cubic polynomial $K_2$ is the Drach integral of motion \cite{ts00}.

\begin{exam} \textbf{The (l) Drach system.}
\par\noindent
Let us consider  another Drach system (case (l) in \cite{ts00}) with the Hamiltonian
\[
H_1=p_xp_y+\alpha \left(y-\dfrac{\rho x}3\right)+\beta x^{-1/2} +\gamma x^{-1/2}(y-\rho x)
\]
Without lost of generality we can put $\rho=-3$. Substituting this Hamiltonian
into the computer programm from \cite{gts05} one gets the separated variables
\[
x=\dfrac{(q_1-q_2)^2}2,\qquad y= \dfrac{(q_1+q_2)^2}{2}
\]
and the  corresponding separated relations
\bq\label{sep-rel-l}
p_j^2=P_j(q_j)= -4\alpha q_j^4\mp 8\sqrt{2}\gamma q_j^3+4H_1q_1^2\mp 4\sqrt{2}\beta q_j+H_2,\qquad j=1,2,
\eq
which give rise a pair of hyperelliptic curves  at $\mu_j=p_j$ and $\lambda_j=q_j$.
Using the St\"ackel matrix
\[
S=\left(\begin{array}{cc}4q_1^2&4q_2^2\\ 1&1\end{array}\right)\,,
\]
we can get the angle variable
\[
\omega_2=\dfrac12\int^{q_1} \dfrac{\mathrm d \lambda}{\sqrt{P_1(\lambda)}}+\dfrac12
\int^{q_2} \dfrac{\mathrm d \lambda}{\sqrt{P_2(\lambda)}},
\]
which is a sum of the incomplete elliptic integrals of the first kind. We do not know how to get algebraic integral of motion starting with this variable $\omega_2$. 

However, we know that there is additional cubic integral of motion
\[
 {K}_3=2(\widetilde{P}'_1\,p_2+\widetilde{P}'_2\,p_1),
\]
which looks like as the Drach integral (\ref{k3-dr}),  but in this case functions
\[\widetilde{P}'_j=\left.(\lambda_1+\lambda_2)^2\dfrac{\partial}{\partial q_j}\, \dfrac{P_j(\lambda_j)}{(\lambda_1+\lambda_2)^4}\right|_{\lambda_{1,2}=q_{1,2}}\,\]
have not simple algebro-geometric explanation. Nevertheless, this polynomial integral $K_3$ has to be a function on polynomials $H_{1,2}$ and elliptic function $\omega_2$. Since, the sum of elliptic integrals
\[
\omega_2=F(H_1,H_2,K_3)
\]
has to be some  function on polynomials $H_{1,2}$ and $K_3$. 

It will be interesting to get this function explicitly and to understand why it exists.

\end{exam}

\begin{exam} \textbf{ The Henon-Heiles system}
\par\noindent
Let us consider the St\"{a}ckel system defined by
\[
S=\left(\begin{array}{cc}1& 1\\ 1& -1\end{array}\right),
\qquad U_{1,2}=a{q_{1,2}^3}\,,
\]
such that
\[
p_{1,2}^2 = H_1\pm H_2+aq_{1,2}^3\,.
\]
The action variables
\bq\label{hh-int}
I_{1}=
H_{1}= \dfrac{p_1^2+p_2^2}{2}-\dfrac{a(q_1^3 + q_2^3)}{2}\,,\qquad
I_{2}=
H_{2}= \dfrac{p_1^2-p_2^2}{2}-\dfrac{a(q_1^3 - q_2^3)}{2}
\eq
have the canonical brackets (\ref{aa-br}) with the angle variables (\ref{w-st}):
\[
w_{1,2}=\dfrac{1}{2}\left( \int^{q_1}\dfrac{\mathrm d \lambda}{\sqrt{a\lambda^3+H_1+H_2}}\pm
\int^{q_2}\dfrac{\mathrm d \lambda}{\sqrt{a\lambda^3+H_1-H_2}}\right)\,.
\]
So, the Hamiltonian system with the hamiltonian $H_1$ has two independent integrals of motion $H_2$ (\ref{hh-int})
and
\bq\label{w-hh}
\omega_2=const\left(
\frac{F\Bigl(\arcsin(z_1),\kappa \Bigr)}{(H_1+H_2)^{1/6}}-
\frac{F\Bigl(\arcsin(z_1),\kappa \Bigr)}{(H_1-H_2)^{1/6}}
\right),
\eq
where $F(z,\kappa)$ is the incomplete elliptic integral of the first kind
and
\[
z_{1,2}=\dfrac{(-1)^{11/12}}{3^{1/4}}\sqrt{1+\frac{a^{1/3} q_{1,2} }{(H_1\pm H_2)^{1/3}}  }\,,\qquad
\kappa=(-1)^{1/3}\,.
\]
After the following change of variables
\[q_{1,2}=\dfrac{x\pm y}{2},\qquad p_{1,2}=p_x\pm p_y \]
the Hamilton function $H_1$ (\ref{hh-int})
\[
H_1= p_x^2+p_y^2-\dfrac{ax(x^2+3y^2)}{8}\,
\]
coincides with one of the Henon-Heiles hamiltonians.

Of course, integral of motion $\omega_2$ (\ref{w-hh}) remains an elliptic function in any coordinates and we can not get single-valued integral of motion on the whole phase space.  
\end{exam}

%%%%%%%%%%%%%%%%%%%%%%%%%%%%%%%%%%%%%%%%%%%%%%%%%%%%%%%%%%%%%%%%%%%%%%%%%%%%%%%%%
%%%%%%%%%%%%%%%%%%%%%%%%%%%%%%%%%%%%%%%%%%%%%%%%%%%%%%%%%%%%%%%%%%%%%%%%%%%%%%%%%%
\section{The Sklyanin algebra}
\setcounter{equation}{0}

In this section we study a class of finite-dimensional Liouville integrable systems described by the representations of the quadratic $r$-matrix Poisson algebra, or the Sklyanin algebra:
\bq
\{\,\on{T}{1}(\lambda),\,\on{T}{2}(\mu)\}= [r(\lambda-\mu),\,
\on{T}{1}(\lambda)\on{T}{2}(\mu)\,]\,, \label{rrpoi}
\eq
Here $\on{T}{1}(\lambda)=T(\lambda)\otimes \mathrm{Id}\,,~\on{T}{2}(\mu)=\mathrm{Id}\otimes T(\mu)$ and
$r(\lambda-\mu)$ is a classical $r$-matrix \cite{skl85}-\cite{skl95}.
In the simplest case
of the $4\times4$ rational $r$-matrix
\bq
r(\lambda-\mu)=\dfrac{\eta}{\lambda-\mu}\Pi,\qquad \Pi=\left(\begin{array}{cccc}
 1 & 0 & 0 & 0 \\
 0 & 0 & 1 & 0 \\
 0 & 1 & 0 & 0 \\
 0 & 0 & 0 & 1
\end{array}\right)\,,\quad \eta\in {\mathbb C}\,,\label{rr}
\eq
 matrix $T(\lambda)$ depends polynomially on
the parameter $\lambda$
\ben\label{22T}
 T(\lambda)&=&\left(\begin{array}{cc}
 A (\lambda)& B (\lambda)\\
 C(\lambda) & D(\lambda)
\end{array}\right)\\
&=&\left(\begin{array}{ll}
\alpha\lambda^n+A_1\lambda^{n-1}+\ldots +A_n\qquad& \beta\lambda^n+B_1\lambda^{n-1}+\ldots+B_n \\
\gamma\lambda^n+ C_1\lambda^{n-1}+\ldots+C_n& \delta\lambda^n+D_1\lambda^{n-1}+\ldots+D_n
\end{array}\right).\nn
\en
The leading coefficients $\alpha,\beta,\gamma,\delta$ and $2n$ coefficients of the $\det T(\lambda)$
\bq
d(\lambda)=\mathrm{det}\,T(\lambda)=(\alpha\delta-\beta\gamma)\lambda^{2n}+Q_1\lambda^{2n-1}+\cdots+Q_{2n}\,.
\label{Acentre}
\eq
are Casimirs of the bracket (\ref{rrpoi}). Therefore, we have a $4n$-dimensional space of the coefficients $A_i, B_i, C_i$ and $D_i$ with $2n$ Casimir operators $Q_i$, leaving us with $n$ degrees of freedom.

\subsection{Open lattices}

For so-called open lattices independent Poisson involutive integrals of motion are given by the coefficients of the entry $A(\lambda)$:
\bq\label{int-open}
A(\lambda) = \alpha\lambda^n +H_1\lambda^{n-1}+\cdots H_n,\qquad \{H_k,H_m\}=0\,.
\eq
In the special action-angle representation \cite{ts07b}, one has $n$ pairs of the action-angle variables:
\bq
A(I_m)=0,\qquad \omega_m=\eta^{-1}\ln B(I_m),\qquad m=1,\ldots,n.
\label{mos-var}
\eq
having the standard Poisson brackets (\ref{aa-br}).

Since the action variables $I_k$ are zeroes of the polynomial
\[A(\lambda)=\alpha\lambda^n +H_1\lambda^{n-1}+\cdots H_n=\alpha\prod_{m=1}^n (\lambda-I_m),\]
initial integrals of motion $H_m$ are elementary symmetric function on $I_m$
\[
H_1=-\sum_{m=1}^n I_m,\quad H_2=\sum_{k\neq m}^n I_kI_m,\quad
%H_3=-\sum_{k\neq l\neq m }^n I_kI_lI_m,
\ldots,\quad H_n=(-1)^n\prod_{m=1}^n I_m
\]
and the matrix $\mathbf J$ (\ref{h-mat}) is equal to
\[
\mathbf J=\left(
  \begin{array}{ccc}
  -1 & \ldots & -1 \\
  \\
  \displaystyle\sum_{m\neq 1}I_m & \ldots & \displaystyle\sum_{m\neq n}I_m \\
  \vdots & \ddots & \vdots \\
  \displaystyle(-1)^n\prod_{m\neq 1}I_m & \cdots &\displaystyle (-1)^n\prod_{m\neq n}I_m \\
  \end{array}
  \right)\,.
\]
In this case all the functions $\phi_k$ (\ref{phi_k}) are functionally independent and one gets that open lattices related with the Sklyanin algebra are maximally superintegrable systems.

We have to underline that functions $\omega_k$ (\ref{mos-var})  may be found without integration, i.e. using pure algebraic constructions. As for the (b) Drach system one gets
\[\{I_i,w_j\}=\{I_i,\ln B(I_j)\}=0,\qquad\Rightarrow\qquad \{I_i,B(I_j)\}=0,\qquad i\neq j.
\]
So, for any Hamiltonian $I_i$ we have additional integrals of motion $B(I_j)$, which are polynomials on momenta and the remaining action variables $I_j$.

\subsection{Periodic lattices}
For so-called periodic lattices integrals of motion are given by the coefficients of the trace of $T(\lambda)$:
\bq\label{int-per}
\tau(\lambda)=\mbox{\rm tr} T(\lambda) = (\alpha+\delta)\lambda^n +H_1\lambda^{n-1}+\cdots H_n,\qquad
 \{H_i,H_j\}=0\,.
 \eq
According to \cite{skl85}-\cite{skl95}, in this case
the separated coordinates $u_i,v_i$ defined by
\[
B(u_i)=0,\qquad v_i= A(u_i)\,,\qquad i=1,\ldots,n,
\]
satisfy to the following separated relations
\bq\label{sep-rel}
 v_i+\det T(u_i)v_i^{-1}=A(u_i)+D(u_i)=\tau(u_i),\qquad i=1,\ldots,n\,.
\eq
The corresponding symplectic form in ($u,v$)-variables is written as
\bq\label{sym-f}
\Omega=\eta\sum_{j=1}^n \mathrm d \log(v_j)\wedge \mathrm du_j\,.
\eq
Let us take the action coordinates  $I_m=H_m$ (\ref{int-per}) instead of $v$-variables \cite{skl95,smirn98}. From (\ref{sep-rel}) one gets
\[
v_j=A(u_j)=\frac12\left( \tau(u_j)+\sqrt {\tau(u_j)-4d(u_j)\,} \right)
\]
and one easily find  expression for the symplectic form in ($I,u$)-variables
\[
\Omega=\eta\sum_{j=1}^n\sum_{k=1}^n \dfrac{u_j^{n-k}}{\sqrt{P(u_j)}}\,\mathrm dI_k\wedge \mathrm du_j\,,
\]
where $P(\lambda)=\tau^2(\lambda)-4d(\lambda)$. The equations of motion
\[
\{\tau(\lambda),u_j\}=\eta\sqrt{P(u_j)\,}\prod_{k\neq j}\dfrac{\lambda-u_k}{u_j-u_k}
\]
are linearized by the Abel transformation \cite{bel97,du81} and the corresponding angle variables look like
\bq\label{w-per}
\omega_j=\eta\sum_{k=1}^n \int^{u_k}\dfrac{ \lambda^{n-j} }{\sqrt{P(\lambda)}}\,\mathrm d\lambda\,,\qquad
\mbox{\rm deg}P(\lambda)=2n.
\eq
In this case matrix $\mathbf J$ (\ref{h-mat}) is identity matrix and integrals of motion $\phi_k=\omega_k$ (\ref{w-per}) are elliptic functions and we could not get single-valued additional integral of motion on the whole phase space.

Recall, that quadratic $r$-matrix algebra (\ref{rrpoi}) is related with a set of
integrable systems that includes $XXX$ Heisenberg magnet, the Toda lattices,
the discrete self-trapping model and the Goryachev-Chaplygin gyrostat \cite{ts07b}.

\begin{exam} \textbf{The Heisenberg magnet}
\par\noindent
Our main model turns out to be an $n$-site Heisenberg magnet, which is an integrable lattice of $n$ sl(2) spins with nearest neighbor interaction.
In the lattice representation the matrix $T(\lambda)$ (\ref{22T}) acquires the following form:
\bq\label{s-mat}
T(\lambda)=L_1(\lambda-c_1)\;L_{2}(\lambda-c_{2})\;\cdots\;L_n(\lambda-c_n)\;
,
\eq
with
\[
L_i(\lambda)=\left(\begin{array}{cc}
u- s_i^{3}& s_i^{+}\\\
s_i^{-}&u+s_i^{3}
\end{array}\right),\qquad i=1,\ldots,n.
\]
Here the local variables $s_i^{\alpha}$, $i=1,\ldots,n$, are generators of $n$ copies
of the sl(2) Poisson algebra:
\bq
\{s^{3}_j,s^{\pm}_j\}=\pm s^{\pm}_j,\qquad
\{s^{+}_j,s^{-}_j\}=2s^{3}_j\,.
\eq
and $c_m$ are arbitrary numbers. If $n=2$ one gets
\[A(\lambda)= \lambda^2+H_1\lambda+H_2=(\lambda-I_1)(\lambda-I_2),\]
where
\[
H_1=-(s_1^{3}+s_2^{3}+c_1+c_2),\qquad
H_2= s_1^3s_2^3+s_1^+s_2^-
+c_1s_2^{3}+c_2s_1^{3}+c_1c_2
\,,\]
and
\[
I_{1,2} = -\dfrac{1}{2}\left(H_1\pm \sqrt{H_1^2-4H_2}\right)\,.
\]
The second polynomial
\[
B(\lambda)=(s^+_1+s^+_2)\lambda-s_1^2s_2^++s_1^+s_2^3-c_1s_2^+-c_2s_1^+
\]
defines the angle variables
\[\omega_{1,2}=\ln B(I_{1,2})\,.\]
The functions $\phi_{1,2}$ (\ref{phi_k})
\[\phi_1= -\dfrac{I_1\omega_1-I_2\omega_2}{I_1-I_2},\qquad
 \phi_2= -\dfrac{\omega_1-\omega_2}{I_1-I_2}\,.
\]
obey to equations $\{H_i,\phi_j\}=\delta_{ij}$. So, the Hamiltonian system with the hamiltonian $H_2$
has two independent integrals of motion $H_1$ and
\[
\Phi=-(I_1-I_2)\,\phi_1=I_1\omega_1-I_2\omega_2=I_1\ln B(I_1)-I_2\ln B(I_2)\,.\nn\\
\]
The similar additional integral for the periodic lattice is some combination of elliptic functions
even in the simplest case $n=2$.

\end{exam}
\begin{exam}\textbf{The Toda lattice}
\par\noindent
The Toda lattices appear as a specialization of our basic model when the parameters are fixed as follows:
\bq
\beta=\gamma=\delta=0\qquad \mbox{\rm and}\qquad \det T(\lambda)=1.
\eq
We also put $\alpha=1$ and $\eta=-1$.
In the lattice representation, the monodromy matrix $T$ (\ref{22T}) acquires
the form
\bq\label{22toda}
T(\lambda)=L_1(\lambda)\cdots L_{n-1}(\lambda)\,L_n(\lambda)
\,, \qquad
L_i=\left(\begin{array}{cc}
 \lambda-p_i &\, -\e^{q_i} \\
 \e^{-q_i}& 0
\end{array}\right)\,.
\eq
Here $p_i,q_i$ are Darboux variables $\{q_i,p_j\}=\delta_{ij}$.
If $n=2$ one gets
\[A(\lambda)= \lambda^2-(p_1+p_2)\lambda+p_1p_2-\e^{q_1-q_2}=(\lambda-I_1)(\lambda-I_2),\]
and
\[
B= -e^{q_2}(\lambda-p_1)
\]
Let us consider natural Hamiltonian systems with integrals of motion
\[
\tilde{H}_1=-H_1=p_1+p_2,\qquad
\tilde{H_2}=\dfrac{p_1^2+p_2^2}{2}+\e^{q_1-q_2}=\dfrac{H_1^2}{2}-H_2.
\]
For these integrals we have
\[
\tilde{\mathbf J}=\left(\begin{array}{cc} 1&1\\I_1&I_2\end{array}\right),\qquad\mbox{and}\qquad
\tilde\phi_1=-\dfrac{I_1\omega_2-I_2\omega_1}{I_1-I_2},\quad\tilde\phi_2=-\dfrac{\omega_1-\omega_2}{I_1-I_2}\,.
\]
So, the Hamiltonian system with the hamiltonian $\tilde{H}_2$ has two independent integrals of motion $\tilde{H_1}$ and
\[
\Phi_1=-(I_1-I_2)\tilde{\phi}_1=I_1\omega_2-I_2\omega_1=-\bigl(I_1\ln B(I_2)-I_2\ln B(I_1)\bigr)\,.
\]
Namely this "generalized angular momentum" has been found in \cite{dam06,deg06}. In contrast with the Drach system we can not directly apply additional theorem in this case. 

At the periodic case one gets
\[
I_1=-(p_1+p_2),\qquad I_2=p_1p_2-\e^{q_1-q_2}-\e^{q_2-q_1},
\]
and the Hamiltonian system with the hamiltonian $H=I_2$ has two independent integrals $I_1$ and
\[
\phi_1=-\dfrac{q_1+q_2}{2}+\dfrac{I_1w_2}{2},\qquad\mbox{where}\qquad
\omega_2=\int^{p_1} \dfrac{\mathrm d\lambda}{\sqrt{(\lambda^2+I_1\lambda+I_2)^2-4\,}}\,.
\]
As for Henon-Heiles system this additional integral of motion
 may be rewritten as a combination of elliptic functions on $p_1$ and $I_{1,2}$.
 
 The global action-angle variables for the  periodic Toda lattices are discussed in  \cite{hk08}. Nevertheless we suppose that periodic Toda lattices are locally maximally superintegrable systems only.
\end{exam}

%%%%%%%%%%%%%%%%%%%%%%%%%%%%%%%%%%%%%%%%%%%%%%%%%%%%%%%%%%%%%%%%%%%%%%%%%%%%%%%%%5
%%%%%%%%%%%%%%%%%%%%%%%%%%%%%%%%%%%%%%%%%%%%%%%%%%%%%%%%%%%%%%%%%%%%%%%%%%%%%%%%%%
\section{Reflection equation algebra}
\setcounter{equation}{0}

In this section we study a class of finite-dimensional Liouville integrable systems described
by the representations of the reflection equation algebra:
\ben
\{\,\on{\mathcal T}{1}(\lambda),\,\on{\mathcal T}{2}(\mu)\}&=& [r(\lambda-\mu),\,
\on{\mathcal T}{1}(\lambda)\on{\mathcal T}{2}(\mu)\,]
\label{ref-poi}\\
&+&
\on{\mathcal T}{1}(\lambda)r(\lambda+\mu)\on{\mathcal T}{2}(\mu)-\on{\mathcal T}{2}(\mu)r(\lambda+\mu)
\on{\mathcal T}{1}(\lambda)
\,. \nn
\en
In the simplest case of the $4\times4$ rational $r$-matrix (\ref{rr})
matrix $\mathcal T(\lambda)$ depends polynomially on
the parameter $\lambda$
\bq\label{22T}
 \mathcal T(\lambda)=\left(\begin{array}{cc}
 \mathcal A(\lambda)& \mathcal B (\lambda)\\
 \mathcal C(\lambda) & \mathcal A(-\lambda)
\end{array}\right), \qquad \mbox{\rm deg} \mathcal T(\lambda)=\left(\begin{array}{cc}
 2n+1& 2n+1\\
 2n-1 & 2n+1
\end{array}\right).
\eq
Coefficients of the entries
\bq
\begin{array}{l}
\mathcal A(\lambda)=\alpha\,\lambda^{2n+1} +\mathcal A_{2n}\,\lambda^{2n}+\mathcal A_{2n-1}\,\lambda^{2n-1}\ldots+\mathcal A_0,\\
\\
\mathcal B(\lambda)=\lambda^{2n+1}
+ \mathcal B_{n}\lambda^{2n-1}+ \mathcal B_{n-1}\lambda^{2n-3} \ldots+ \mathcal B_1\lambda,\\
\\
\mathcal C(\lambda)=\mathcal C_{n}\lambda^{2n-1}+\ldots+\mathcal C_{2}\lambda^{3}+\mathcal C_1\lambda,
\end{array}
\label{asymp1}
\eq
are generators of the quadratic Poisson algebra (\ref{rrpoi}).
The leading coefficient $\alpha$ and $2n+1$ coefficients of the $\det T(\lambda)$
\bq
d(\lambda)=\mathrm{det}\,\mathcal T(\lambda)=Q_{2n}\lambda^{4n}+Q_{2n-1}\lambda^{4n-2}+\cdots+Q_{0}\,.
\label{Acentre}
\eq
are Casimirs of the bracket (\ref{rrpoi}). Therefore, we have a $4n+1$-dimensional space of the coefficients
\bq
\mathcal A_0,\ldots, \mathcal A_{2n},~\mathcal B_1,\ldots, \ \mathcal B_{n},~ \
\mathcal C_1,\ldots, \ \mathcal C_{n} \label{var1}
\eq
with $2n+1$ Casimir operators $Q_i$, leaving us with $n$ degrees of freedom.

\subsection{Open lattices}
For so-called open lattices integrals of motion are given by the coefficients of the entry $\mathcal B(\lambda)$:
\bq\label{int-open-ref}
\mathcal B(\lambda) = \lambda^{2n+1}+ H_{1}\lambda^{2n-1}+ H_{2}\lambda^{2n-3} \ldots+ H_n\lambda,\qquad \{H_k,H_m\}=0\,.
\eq
In the special action-angle representation \cite{ts07c}, one has $n$ pairs of the action-angle variables:
\bq
\mathcal B(\pm I_m)=0,\qquad \omega_m=\eta^{-1}\ln \mathcal A(I_m),\qquad m=1,\ldots,n,
\label{dn-var}
\eq
having the standard Poisson brackets (\ref{aa-br}).

As above the action variables $I_k$ are zeroes of the polynomial
\[\mathcal B(\lambda)=\lambda^{2n+1}+ H_{1}\lambda^{2n-1}+ H_{2}\lambda^{2n-3} \ldots+ H_n\lambda=\lambda\prod_{m=1}^n (\lambda^2-I_m^2),\]
integrals of motion $H_m$ are elementary symmetric function on $I_m^2$
\[
H_1=-\sum_{m=1}^n I_m^2,\quad H_2=\sum_{k\neq m}^n I_k^2I_m^2,\quad
%H_3=-\sum_{k\neq l\neq m }^n I_kI_lI_m,
\ldots,\quad H_n=(-1)^n\prod_{m=1}^n I_m^2
\]
and the matrix $\mathbf J$ (\ref{h-mat}) is equal to
\[
\mathbf J=2\left(
  \begin{array}{ccc}\displaystyle
  -I_1 & \ldots &\displaystyle -I_n \\
  \\
  \displaystyle I_1\sum_{m\neq 1}I_m^2 & \ldots & \displaystyle I_n\sum_{m\neq n}I_m^2 \\
    \vdots & \ddots & \vdots \\
  \displaystyle(-1)^nI_1\prod_{m\neq 1}I_m^2 & \cdots &\displaystyle (-1)^nI_1\prod_{m\neq n}I_m^2 \\
  \end{array}
  \right)\,.
\]
As above all the functions $\phi_j$ (\ref{phi_k}) are functionally independent and one gets that open lattices related with the reflection equation algebra are maximally superintegrable systems.

We have to underline that functions $\omega_k$ (\ref{dn-var}) and, therefore, integrals of motion $\phi_k$ may be found without integration, i.e. using pure algebraic constructions. As above any action variables $I_i$ have additional polynomial integrals of motion $\mathcal A(I_j)$
\[\{I_i,w_j\}=\{I_i,\ln \mathcal A(I_j)\}=0,\qquad\Rightarrow\qquad \{I_i,\mathcal A(I_j)\}=0,\qquad i\neq j,
\]
but for the integrals $H_k$ one gets some combinations of logarithms and $I_j$ only.

\subsection{Periodic lattices}
The theory of periodic lattices is based on the following construction of commutative subalgebras \cite{skl88,sokts}.
Let us introduce the boundary matrix
\bq
 \mathcal K(\lambda)=\left(\begin{array}{cc}
 {a}(\lambda) & 0 \\
{c}(\lambda) & d(\lambda)
\end{array}\right)\,\label{KK}
\eq
whose entries $ a(\lambda),d(\lambda)$ are polynomials with numerical coefficients and entry $ c(\lambda)$ is arbitrary polynomial on $\lambda$.
If the polynomial
\[\tau(\lambda)=\mbox{\rm tr}\,\mathcal K(\lambda) \mathcal T(\lambda)=
a(\lambda)\mathcal A(\lambda)+c(\lambda)\mathcal B(\lambda)+d(\lambda)\mathcal A(-\lambda)
\]
has $n$ independent dynamical coefficients $H_1,\ldots,H_n$ only, then
\[
\{\tau(\lambda),\tau(\mu)\}=0,\qquad\Rightarrow\qquad \{H_i,H_j\}=0,\qquad i,j=1,\ldots,n.
\]
These Poisson involutive integrals of motion $H_i$ define the Liouville integrable systems or the periodic lattices related with the reflection equation algebra.

In the periodic case the $n$ pairs of the former action-angle variables (\ref{dn-var}) for the open lattices
\bq
\mathcal B(\pm u_m)=0,\qquad v_m=\mathcal A(u_m),\qquad m=1,\ldots,n,
\label{dn-var-per}
\eq
are the simple separated variables, which
satisfy to the following separated relations
\bq\label{sep-ref}
a(u_j)v_j+d(u_j)\det\mathcal T(u_j)\,v_j^{-1}=\tau(u_j)\,.
\eq
The corresponding symplectic form in ($u,v$)-variables has the form (\ref{sym-f}).
Therefore, we can apply the above consideration of periodic lattices related with the Sklyanin algebra to construction of the angle variables (\ref{w-per}) in this case too.

As above the action variables $I_m$ coincide with integrals of motion $H_m$ and the matrix $\mathbf J$ (\ref{h-mat}) is identity matrix. Additional integrals of motion $\phi_k$ coincide with the angle variables $\omega_m$ (\ref{w-per}), which  are integrals of Abelian differentials on the hyperelliptic curve $\mu^2=P(\lambda)$ defined by (\ref{sep-ref}), i.e. they are elliptic functions. So, all the periodic lattices related with the reflection equation algebra are formally maximally superintegrable systems, i.e. locally superintegrable only.

Recall, that reflection equation algebra (\ref{ref-poi}) is related with a set of
integrable systems that includes the generalized Toda lattices,
the Kowalevski top and $XXX$ Heisenberg magnet with boundary conditions \cite{ts07c}.

\begin{exam}\textbf{Open Toda lattice}
\par\noindent
According to \cite{skl88,sokts} the $2\times 2$ Lax matrix for the generalized open Toda lattice
\bq\label{toda22}
\mathcal T(\lambda)=\left(\prod_{k=1}^n L_k(\lambda)\right)\mathcal K_-(\lambda)
\left( \prod_{k=1}^n L_k(-\lambda)\right)^{-1}
 \,,
\eq
where $L_i$ is given by (\ref{22toda}) and
\[
\mathcal K_-(\lambda)=\left(
 \begin{array}{cc}
 2a_2\lambda^2-ia_1\lambda+a_0 & (4a_2\e^{q_n}-1)\lambda \\
 0 & 2a_2\lambda^2+ia_1\lambda+a_0
 \end{array}
 \right),\]
satisfies to the reflection equation algebra at $\alpha=0$ and $\eta=1$.
Here $p_i,q_i$ are dynamical variables and $a_k$ are parameters.

All the open Toda lattices associated with classical root systems
 $\mathscr{B}_n$, $\mathscr{C}_n$, $\mathscr{BC}_n$ and $\mathscr{D}_n$
are isomorphic to each other \cite{ts04} and, therefore, we consider $\mathscr{B}_n$ root system only.
In this case $a_2=a_1=0$ and at $n=2$ one gets
{\setlength\arraycolsep{1pt}
\ben
\mathcal B&=&\lambda^5+H_1\lambda^3+H_2\lambda=\lambda(\lambda-I_1^2)(\lambda-I_2^2)=\nn\\
\nn\\
&=&\lambda^5-(p_1^2+p_2^2-2\e^{q_1-q_2}-2a_0\e^{q_2})\lambda^3
+\bigl( (p_1p_2-\e^{q_1-q_2})^2-2a_0(p_1^2\e^{q_2}+\e^{q_1})\bigr)\lambda,\nn
\en}
such that
\[
H_1= -I_1^2-I_2^2,\quad H_2=I_1^2I_2^2,\qquad
\phi_1=-\dfrac{I_1\omega_1-I_2\omega_2}{2(I_1^2-I_2^2)},\quad
\phi_2=\dfrac{I_1\omega_2-I_2\omega_1}{2I_1I_2(I_1^2-I_2^2)}\,.
\]
So, the Hamiltonian system with the hamiltonian $H_1$ has two independent integrals of motion $H_2$ and
\[
\Phi_2=2I_1I_2(I_1^2-I_2^2)\phi_2=I_1\omega_2-I_2\omega_1=-\bigl(I_1\ln \mathcal A(I_2)-I_2\ln \mathcal A(I_2)\bigr),
\]
where
\[
\mathcal A=-\e^{-q_1}\Bigl(
\lambda^4-p_1\lambda^3-(p_2^2+2a_0\e^{q_2}+\e^{q_1-q_2})\lambda^2
+(p_1p_2^2-p_2\e^{q_1-q_2}+2a_0p_1\e^{q_2})\lambda\Bigl)-a_0\,.
\]
Such as matrix $\mathbf J$ (\ref{h-mat}) has the special form, the additional integral of motion $\Phi_2$ has the form of the angular momentum in ($I,\omega$)-variables (see \cite{ts00} and \cite{deg06}). However at $n>2$ we will have completely another picture.
\end{exam}

\section{Conclusion}
We discuss some constructions of the single valued integrals of motion on the whole phase space by using multi-valued action-angle variables.

For the St\"ackel systems we use addition theorem for the construction of the polynomial additional integrals of motion starting with  zero-genus hyperelliptic curves. The construction of the polynomial additional integrals of motion is an algebraic procedure for the open lattices related with the Sklyanin algebra or with the reflection equation algebra. On the other hand  for the periodic lattices one gets the sums of integrals of first kind Abelian differentials on the hyperelliptic curves, which is locally defined only.

It will be interesting to consider inverse problem and try to find algorithm of construction
action-angle variables starting with known additional polynomial integrals of motion.

 The research was partially supported by
the RFBR grant 06-01-00140.

\end{document}